\documentclass[aps,pre,twocolumn,groupedaddress]{revtex4}
\usepackage{graphicx}
\usepackage{dcolumn}
\usepackage{bm}
\usepackage{amssymb}
\usepackage{color}

\begin{document}

\title{Construction of nuclear envelope shape by a high-genus vesicle with pore-size constraint}

\author{Hiroshi Noguchi}
\affiliation{Institute for Solid State Physics, University of Tokyo,
 Kashiwa, Chiba 277-8581, Japan}

\begin{abstract}
Nuclear pores have an approximately uniform distribution in the nuclear envelope of most living cells.
Hence, the morphology of the nuclear envelope is a spherical stomatocyte with a high genus.
We have investigated the morphology of high-genus vesicles under pore-size constraint using dynamically triangulated membrane simulations.
Bending-energy minimization without volume or other constraints
produces a circular-cage stomatocyte, where the pores are aligned in a circular line on an oblate bud.
As the pore radius is reduced, the circular-pore alignment is more stabilized than a random pore distribution on a spherical bud.
However, we have clarified the conditions for the formation of a spherical stomatocyte:
a small perinuclear volume, osmotic pressure within nucleoplasm, and repulsion between the pores.
When area-difference elasticity is taken into account,
the formation of cylindrical or budded tubules from the stomatocyte and discoidal stomatocyte is found.
\end{abstract}

\maketitle

\section{Introduction}

The nucleus of a eukaryotic cell is surrounded by a nuclear envelope.
The nuclear envelope consists of two bilayer membranes
 connected by many lipidic pores, which are supported by 
a protein complex called the nuclear pore complex (NPC) \cite{fabr97,webs09,hetz10,hoel11,raic12,gros12}.
The NPC is an aqueous channel that shows an eight-fold rotational symmetry with an outer diameter of 100 $\sim$ 120 nm
and a central transport channel with a diameter of $\sim$ 40 nm.
The outer  nuclear membrane is also connected  to the endoplasmic reticulum (ER),
which consists of tubular networks and flat membranes.
The inner nuclear membrane contains proteins that interact with 
nuclear lamina, chromatin-associated proteins, and other nuclear proteins.
The nuclear lamina is a network of lamin filaments and associated proteins underlying the inner membrane.

Both the size of the nucleus and NPC density in the nuclear envelope vary among different organisms \cite{fabr97,gros12,maul77,jorg07}.
The average number of NPCs in the nuclear envelope of
vertebrates is typically 2000--5000 (10--20 pores $\mu$m$^{-2}$). However, in a single Xenopus laevis oocyte nucleus,
$5 \times 10^7$ NPCs (60 pores $\mu$m$^{-2}$) are found. 
Yeast have barely 60--200 NPCs per nucleus (10--20 pores $\mu$m$^{-2}$) \cite{wine97}.
Depletion of nucleoporins such as NSP1 and NIC96 induces a significantly decreased NPC density \cite{fabr97}.
The number of the nuclear pores and the area of the nuclear envelope are doubled during interphase \cite{hetz10,wine97,maes10}.
For most metazoan cells,
the nuclear envelope is disassembled during early mitosis and reassembled to form two daughter nuclei.
In contrast, most fungi cells undergo closed mitosis, in which the nuclear envelope does not disassemble but deforms
into a dumbbell shape  and is subsequently divided into two \cite{webs09,hetz10,wine97}.
Thus, the nuclear envelope dynamically changes its shape during mitosis.

During interphase, the nuclei of most cells are spherical or ellipsoidal
and the NPCs have an approximately uniform distribution.
However, blebbed or lobed nuclei are observed in various diseases as well as aging \cite{webs09,jevt14,funk13}.
Canine parvovirus infection induces the accumulation of the NPCs and lamin B1
in the apical side,  where new virus capsides are formed \cite{mant15}.
Whether the nuclear size and shape directly impact chromatin organization and gene expression remains an open question, but
some correlations have been reported  \cite{webs09,jevt14}.

In this paper, we consider a high-genus vesicle as a basic model system for the nuclear envelope.
If the ER is removed, the morphology of the nuclear envelope is a spherical stomatocyte of the genus $g\gg 0$,
which can be formed by a single-component membrane.
Theoretically, it can be treated as a closed curved sheet that is incompressible but has no shear elasticity (in a fluid phase).

The morphology of genus-0 vesicles has been intensively investigated, both
experimentally and theoretically~\cite{canh70,helf73,seif97,hota99,lipo14,svet14,tu14,saka12,kahr12a,saka14}.
Discocyte, prolate, and stomatocyte,
can be reproduced by the minimization of bending energy with area and volume constraints.
Other shapes, such as pear and branched tubes, are obtained by the addition of 
 spontaneous curvature or area-difference elasticity (ADE)~\cite{seif97,svet14}.

Compared to the genus-0 vesicles,
vesicle shapes with $g>0$ have been much less explored.
Besides, most of the previous papers on vesicles with $g>1$ focused on vesicle shapes 
with $g=1$ and $2$ \cite{zhon90,seif91a,four92,juli93,juli93a,juli96,mich94,mich95,nogu15a,bouz15}.
To our knowledge, higher-genus vesicles have been studied only in two studies, including our previous one \cite{halu02,nogu15c}.
The budding of hexagonally arranged pores in polymersomes was investigated experimentally and theoretically in Ref.~\cite{halu02}.
Vesicle shapes with $2\le g \le 8$ were investigated using dynamically triangulated membrane simulations in our previous study \cite{nogu15c}.
Bending-energy minimum states in the absence of volume and other constraints are
circular-cage stomatocytes, where $g+1$ pores are aligned along the circular edge of an inner bud.
With the volume constraint,
the reduction of the vesicle volume results in the formation of a spherical stomatocyte,
where the pores are distributed on a spherical bud.
This change in the pore arrangement alters the characteristics of the shape transition for $g \ge 3$.
With an increasing intrinsic area difference,  
the circular-cage  stomatocyte continuously transforms into a discocyte with a line of $g$ pores.
In contrast, at small volumes, a spherical stomatocyte transforms into a ($g+1$)-hedral shape 
and subsequently exhibits a discrete phase transition to a discocyte.

In this study,
we investigated the conditions for the formation of a spherical stomatocyte with small pores.
We considered a liposome, in which NPCs are embedded.
Since the NPC fixes the pore size in the nuclear envelope,
we constrained the maximum pore size by a toroidal ring.
The inside of the vesicle corresponds to the perinuclear space between the outer and inner nuclear membranes.
For the nucleus, the inner bud is the nucleoplasm space that contains nucleosomes and nuclear lamina.
Since the transport through the nuclear pores is regulated by the NPCs,
the composition of the nucleoplasm is different from outside of the nucleus, cytoplasm.
As mentioned above, the small vesicle volume (perinuclear volume) is a sufficient condition, but 
the volume is not a good control parameter for the nucleus since its volume is shared with the ER.
The ER widely spreads in the cell and its shape and size dynamically change.
Hence, we also surveyed other constraints or interactions:
the osmotic pressure between the nucleoplasm and cytoplasm, repulsion between the pores, and the intrinsic area difference.
We discuss the conditions to stabilize the spherical shape of the nuclear envelope based on our simulation results.

\section{Simulation Model and Method}\label{sec:method}

Fluid vesicles are simulated by 
 a dynamically triangulated surface method \cite{gomp04c,nogu09,nogu15a,nogu15c}.
Since the details of the potentials are described in Ref.~\cite{nogu15a} and
the general features of the triangulated membrane can be found in Ref.~\cite{gomp04c},
 we briefly describe the membrane model here.
A vesicle consists of
$4000$ vertices with a hard-core excluded volume of diameter $\sigma_0$. 
The maximum bond length is $\sigma_1=1.67\sigma_0$.
The volume $V$ of the perinuclear space and membrane surface area $A$ are maintained by harmonic potentials $U_{\rm {V}}= (1/2)k_{\rm {V}}(V-V_{\rm 0})^2$ and
$U_{\rm {A}}= (1/2)k_{\rm {A}}(A-A_{\rm 0})^2$ with  $k_{\rm {V}}=4k_{\rm B}T$ and $k_{\rm {A}}=8k_{\rm B}T$,
where $k_{\rm B}T$ is the thermal energy.
The deviations in the area $A$, the volume $V$, and reduced volume $V^*=V/(4\pi/3){R_{\rm {ves}}}^{3}$ from the target values are less than $0.1$\%,
where $R_{\rm {ves}}=\sqrt{A/4\pi}$.

The bending energy of a homogeneous fluid vesicle is given by~\cite{canh70,helf73}
\begin{equation}
U_{\rm {cv}} =  \int  \frac{\kappa}{2}(C_1+C_2)^2   dA,
\label{eq:cv}
\end{equation}
where $C_1$ and $C_2$ are the principal curvatures at each point 
in the membrane. The coefficient $\kappa$ is the bending rigidity.
The spontaneous curvature and Gaussian bending energy
are not taken into account, since the spontaneous curvature
 vanishes for a membrane whose inner and outer monolayers consist of the same lipid compositions,
and the integral over the Gaussian curvature $C_1C_2$ is invariant for a fixed topology.

In the ADE model, the ADE energy $U_{\rm {ADE}}$ is added as follows \cite{seif97,svet14}:
\begin{equation}
U_{\rm {ADE}} =  \frac{\pi k_{\rm {ade}}}{2Ah^2}(\Delta A - \Delta A_0)^2.
\label{eq:ade}
\end{equation}
The areas of the outer and inner monolayers of a bilayer vesicle
differ by $\Delta A= h \int (C_1+C_2) dA$,
where $h$ ($\simeq 2$ nm) is the distance between the two monolayers.
The area differences are normalized by a spherical vesicle as
$\Delta a =\Delta A/8\pi h R_{\rm {ves}}$ and $\Delta a_0 = \Delta A_0/8\pi h R_{\rm {ves}}$
 to show our results.
The spherical vesicle with $\Delta a_0=0$ has $\Delta a =1$ and $U_{\rm {ADE}} =8\pi^2 k_{\rm {ade}}$.
The mean curvature at each vertex is discretized using dual lattices \cite{nogu15a,gomp04c,itzy86,nogu05}:
\begin{equation}
(C_1+C_2){\bf n}_i = \frac{1}{\lambda_i}  
  \sum_{j(i)} \frac{\lambda_{i,j}{\bf r}_{i,j}}{r_{i,j}},
\end{equation}
where the sum over $j(i)$ is for the neighbors of the $i$-th vertex, which are
connected by bonds. The bond vector between the vertices $i$ and $j$ is
${\bf r}_{i,j}$, and $r_{i,j}=|{\bf r}_{i,j}|$. 
The length of a bond in the dual lattice is 
$\lambda_{i,j}=r_{i,j}[\cot(\theta_1)+\cot(\theta_2)]/2$.
The angles $\theta_1$ and $\theta_2$ are opposite to bond $ij$ in 
the two triangles sharing this bond,
  and $\lambda_i=0.25\sum_{j(i)} \lambda_{i,j}r_{i,j}$ is the area of the dual cell.
The normal vector ${\bf n}_i$ points from inside of the vesicle to outside.

\begin{figure}
\includegraphics{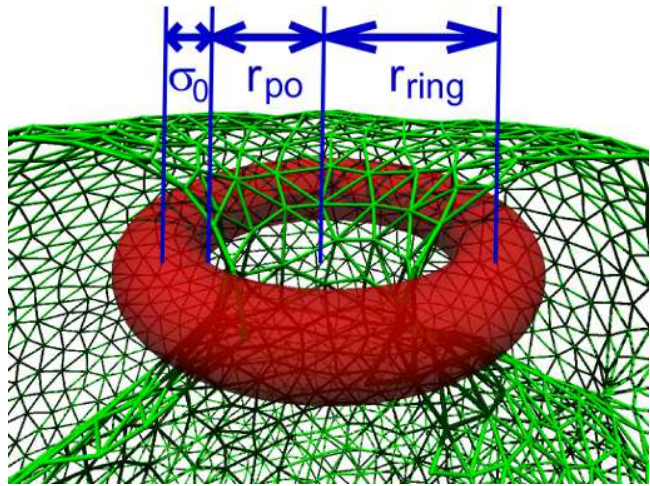}
\caption{
Ring to constrain the pore size.
The membrane and ring are displayed in light gray (green) and in dark gray (red), respectively.
The front half of the membrane is removed for clarity.
}
\label{fig:ring}
\end{figure}

In a nuclear envelope, the NPC fixes the pore radius by binding it from inside.
However, if the positions of membrane vertices are directly constrained by an attractive potential,
the vertex diffusion on membrane is intensively suppressed.
Therefore, instead we add a ring to surround the pore and this ring determines the maximum pore radius.
In this way, it does not restrict membrane diffusion through the pore so that the ratio of inner and outer membranes can relax into 
their equilibrium values.
For a genus-$g$ vesicle, $g+1$ rings are required
since the stomatocyte has  $g+1$ pores.
The ring has hard-core excluded volume interaction with membrane vertices 
as a torus with a major radius $r_{\rm {ring}}$ and minor radius $\sigma_0$ as shown in Fig.~\ref{fig:ring}. The membrane vertices are forbidden to enter the inside of the ring.
Hence, the maximum pore radius is $r_{\rm {po}}=r_{\rm {ring}}-\sigma_0$.
This minor radius is sufficiently large to prevent the membrane penetration.
In our previous study~\cite{saka14}, we considered a genus-0 vesicle constrained in a sphere.
Here, the membrane pores are constrained by rings that can move and rotate.

To generate a positive osmotic pressure  $P_{\rm {in}}$ between nucleoplasm and cytoplasm,
$n_{\rm {in}}$ particles are added in the nucleoplasm.
These inner particles have a hard-core excluded volume interaction with membrane vertices for a diameter of $5.3\sigma_0$
to avoid penetration through the membrane pores.
To obtain an ideal gas, the particles have no interaction with each other.

After equilibration,
the pressure $P_{\rm {in}}$ is calculated
from the equation of the state of an ideal gas, 
$P_{\rm {in}}=n_{\rm {in}}k_{\rm B}T/V_{\rm {in}}$.
The volume $V_{\rm {in}}$ for the inner particles to move
is calculated from 150 snapshots for each condition as follows.
For each membrane conformation,
the particles are moved by Monte Carlo (MC) methods
and 4,000,000$n_{\rm {in}}$ positions are obtained.
The rectangular space to cover these positions is divided into $80\times 80\times 80$ boxes
and the number of boxes containing particles is counted.

To investigate the interactions between the pores,
a hard-core excluded volume interaction is added between the ring centers and
the hard-core radius $r_{\rm {ex}}$ is varied.
Unless otherwise specified, no inner particles ($n_{\rm {in}}=0$) and no ring--ring interactions ($r_{\rm {ex}}=0$) are employed
so that the rings can be overlapped.

In the present simulations,
we use  $\kappa=20k_{\rm B}T$ and $k_{\rm {ade}}^* = k_{\rm {ade}}/\kappa=1$.
These are typical values for phospholipids~\cite{seif97,saka12}.
Most of the simulations are performed with pore-constraint rings 
in the absence of the ADE potentials under the volume and area constraints.
In some of the simulations,  $k_{\rm {V}}=0$ or the ADE potential is employed 
to simulate the vesicles without volume constraints or with the ADE model, respectively.
A Metropolis MC method is used for the motion of membrane vertices and rings
and the reconnection of the bonds (bond flip).
The canonical MC simulations are performed 
from different initial conformations for various conditions.
To obtain the free-energy profile for $V^*$
a replica exchange MC method \cite{huku96,okam04} with $16$ replicas is employed for the genus-5 vesicles
for $n_{\rm {in}}=0$, $6$, and $10$ at $r_{\rm {po}}/R_{\rm {ves}}= 0.2$
and for $r_{\rm {po}}/R_{\rm {ves}}= 0.14$ and $n_{\rm {in}}=0$.
A value of the free energy $F$ at $V^*=0.54$ is taken to be the origin.
Error bars are calculated from three or four independent runs.
The vesicle radius $R_{\rm {ves}}$ is employed as the length unit in this paper.
Since $r_{\rm {po}} \simeq 50$ nm for the nuclear pores,
the simulated vesicle size is $R_{\rm {ves}} \simeq 650$ nm for $r_{\rm {po}}/R_{\rm {ves}}= 0.077$. This corresponds to a nucleus of small yeasts \cite{gros12,jorg07}.

Before showing the simulation results,
we discuss the relation between the reduced volume $V^*$ and the distance $d$ between the outer and inner nuclear membranes
to map the value of $V^*$ to the nuclear shape.
When the outer and inner membranes are assumed to be spheres with radii $R_1$ and $R_2$,
the reduced volume is given by $V^*= (R_1^3-R_2^3)/(R_1^2+R_2^2)^{3/2}$.
Hence, for $V^*\ll 1$, $V^*= 3d/2R_{\rm {ves}}$ where the average distance $d= R_1-R_2$.
Thus, the distance between two membranes is proportional to $V^*$ for small $V^*$.

\section{Results}\label{sec:results}

\subsection{Genus-5 vesicles}\label{sec:g5}

\begin{figure}
\includegraphics{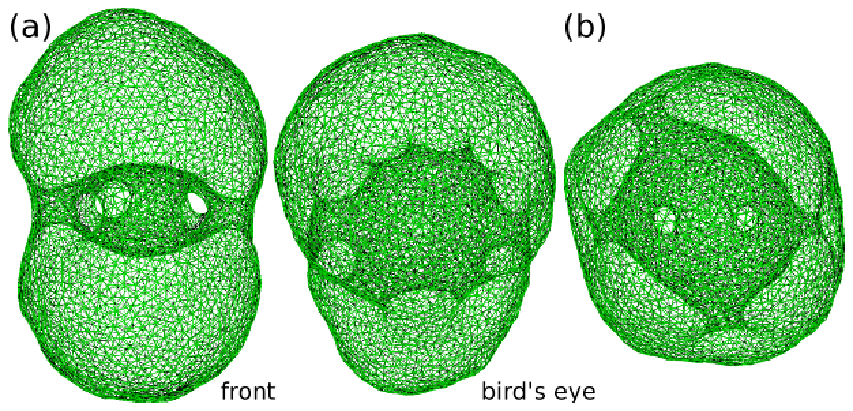}
\includegraphics{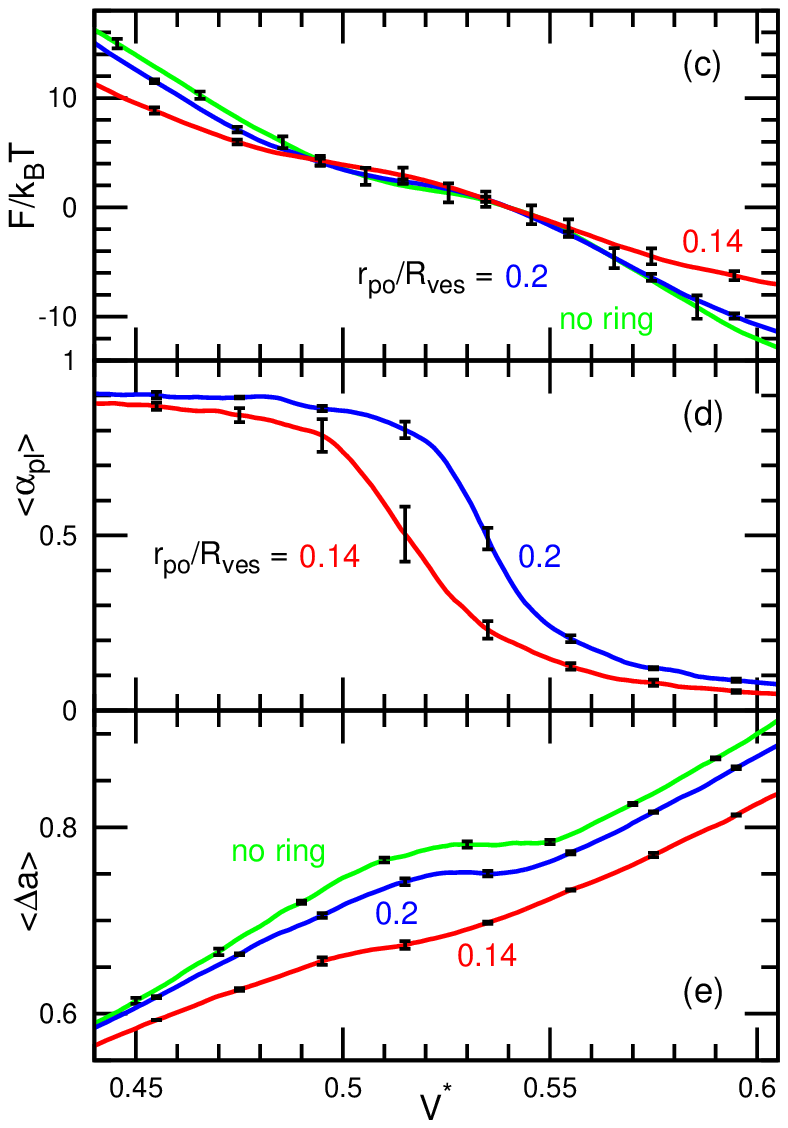}
\caption{
Comparison of genus-5 vesicles without pore-size constraint (no ring) 
and  with six rings of $r_{\rm {po}}/R_{\rm {ves}}= 0.14$ and $0.2$.
(a),(b) Snapshots at (a) $V^*=0.6$ and (b) $V^*=0.45$  with no ring. 
Front and bird's eye views are shown in (a).
(c)--(e) Dependence on the reduced volume $V^*$.
(c) Free-energy profile $F$. (d) Mean aplanarity $\langle\alpha_{\rm {pl}}\rangle$.
(e) Mean area difference $\langle\Delta a\rangle$.
Error bars are shown at several data points.
}
\label{fig:fr}
\end{figure}

First, we show the morphology of genus-5 vesicles without the ADE energy.
In our previous paper \cite{nogu15c}, we reported 
that without the pore-size constraint, the vesicles exhibit a circular-cage stomatocyte 
and spherical stomatocyte shape for $V^* \gtrsim 0.55$ and $V^* \lesssim 0.52$, respectively. 
In the circular-cage stomatocyte, 
its inner bud has an oblate shape and six pores are aligned along the circular edge of the oblate 
[see Fig.~\ref{fig:fr}(a)].
In the spherical stomatocyte, six pores are distributed on the spherical surface
[see Fig.~\ref{fig:fr}(b)].
Around the transient volume, $0.52 \lesssim V^* \lesssim 0.55$, 
the distribution of the pores fluctuates around the circular positions.
With decreasing $V^*$, the positions  more frequently deviate from a plane.
In the absence of the volume constraint, 
the vesicle forms the circular-cage stomatocyte around $V^*\simeq 0.63$ and $\Delta a\simeq 1$
and the vesicle free-energy $F$ increases with decreasing $V^*$ \cite{nogu15c} [see Fig.~\ref{fig:fr}(c)].

\begin{figure}
\includegraphics{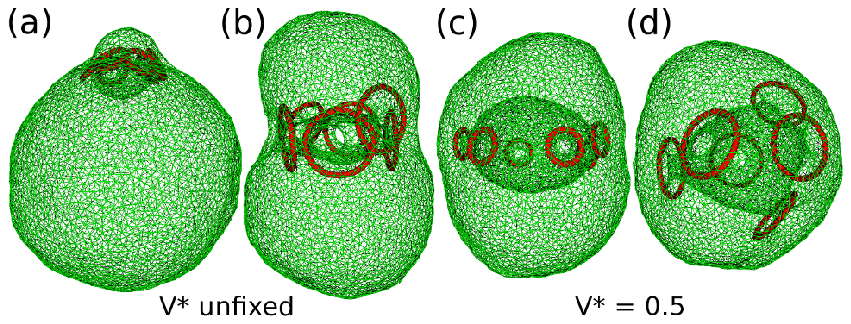}
\includegraphics{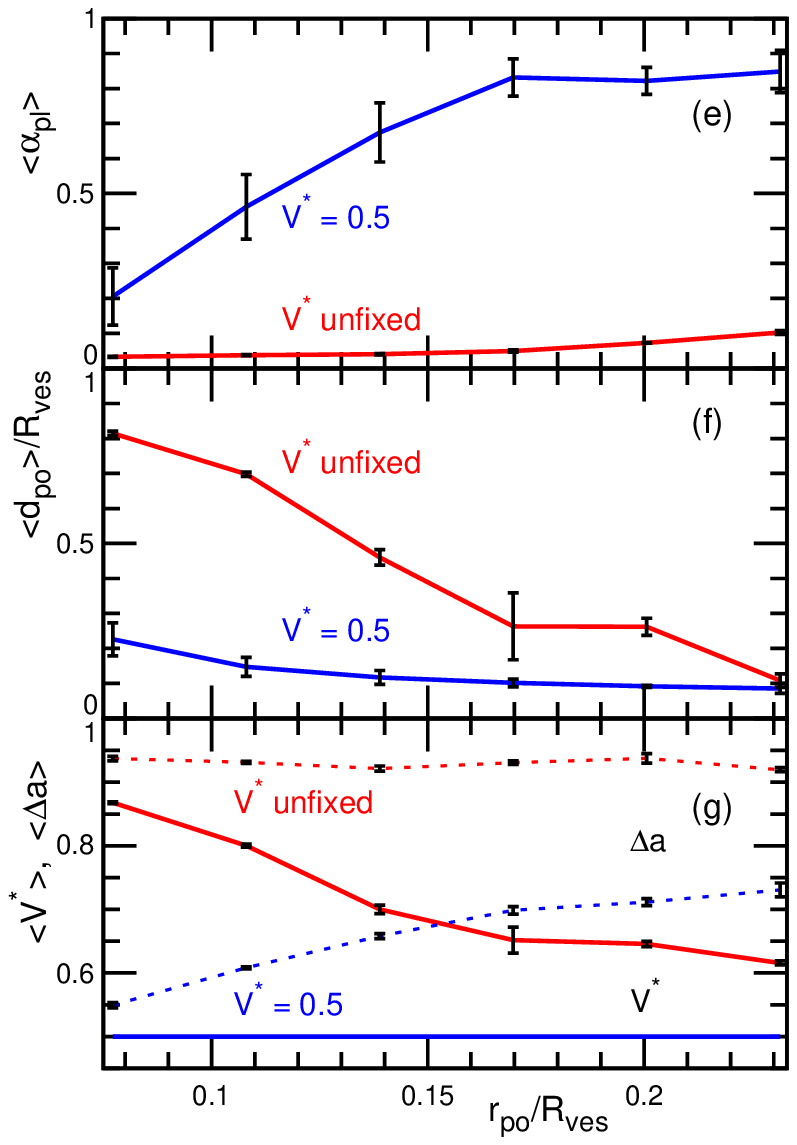}
\caption{
Dependence on the maximum pore radius $r_{\rm {po}}$ for genus-5 vesicles
with or without the volume constraint.
(a)--(d) Snapshots at (a),(c) $r_{\rm {po}}/R_{\rm {ves}}=0.077$ 
and (b),(d) $r_{\rm {po}}/R_{\rm {ves}}=0.2$.
The rings are depicted with a half of the real minor raidus for clarity.
(e)--(g)  Dependence of the averages of (e) aplanarity $\langle\alpha_{\rm {pl}}\rangle$,
(f) distance $\langle d_{\rm {po}}\rangle$ between 
centers of the rings and vesicle, 
(g) reduced volume $\langle V^*\rangle$, and area difference $\langle\Delta a\rangle$.
The solid and dashed lines in (g) represent $\langle V^*\rangle$ and 
$\langle\Delta a\rangle$, respectively.
}
\label{fig:rr}
\end{figure}

When the pore size is constrained by the toroidal rings as shown in red in Figs.~\ref{fig:rr}(a)--(d),
the vesicle shape is modified.
As the radius $r_{\rm {po}}$ is reduced, the transition volume $V^*$ from circular-cage to spherical stomatocytes
decreases.
This transition can be characterized by a change in a shape parameter, aplanarity $\alpha_{\rm {pl}}$, 
of the center of the rings [see Fig.~\ref{fig:fr}(d)]. 
It is defined as \cite{nogu06,nogu12a}
\begin{equation}
\alpha_{\rm {pl}} = \frac{9\lambda_1\lambda_2\lambda_3} {(\lambda_1+\lambda_2+\lambda_3)
    (\lambda_1\lambda_2+\lambda_2\lambda_3+\lambda_3\lambda_1)},
\end{equation}
where ${\lambda_1}\leq {\lambda_2}\leq {\lambda_3}$ are three eigenvalues
of the gyration tensor
$a_{\alpha\beta}= (1/n_{\rm {ring}})\sum_i (\alpha_{i}-\alpha_{\rm G})
(\beta_{i}-\beta_{\rm G})$
with $\alpha, \beta \in x,y,z$ and six rings ($n_{\rm {ring}}=6$).
The center of mass of $n_{\rm {ring}}$ rings is given by $\alpha_{\rm G} = (1/n_{\rm {ring}})\sum_i \alpha_{i}$.
The aplanarity $\alpha_{\rm {pl}}$ quantifies the deviation from a planar shape 
and takes minimum $0$ and maximum $1$ for a plane ($\lambda_1=0$) 
and sphere ($\lambda_1=\lambda_2=\lambda_3$), respectively.
Hence, the circular-cage and spherical stomatocytes have  $\alpha_{\rm {pl}}\simeq 0.1$ and $0.9$, respectively.
The area difference $\Delta a$ exhibits an S-shape at the transition,
but it becomes obscure for smaller $r_{\rm {po}}$ [see Fig.~\ref{fig:fr}(e)].
When the volume is fixed to $V^*=0.5$, 
the vesicle transforms to the circular-cage shape
with a decrease in $r_{\rm {po}}$ [see Figs.~\ref{fig:rr}(c)--(e)].
Thus, the pore-size constraint stabilizes the circular-cage shape
and a smaller $V^*$ is required to form the spherical stomatocyte.

\begin{figure}
\includegraphics{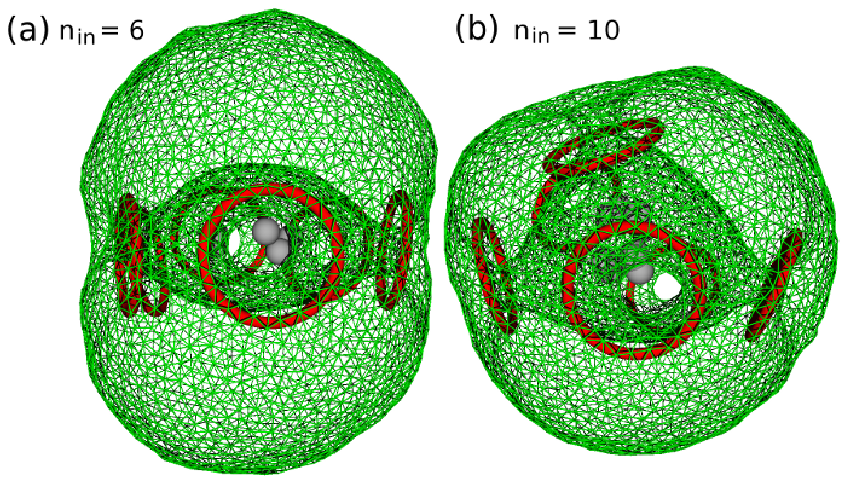}
\includegraphics{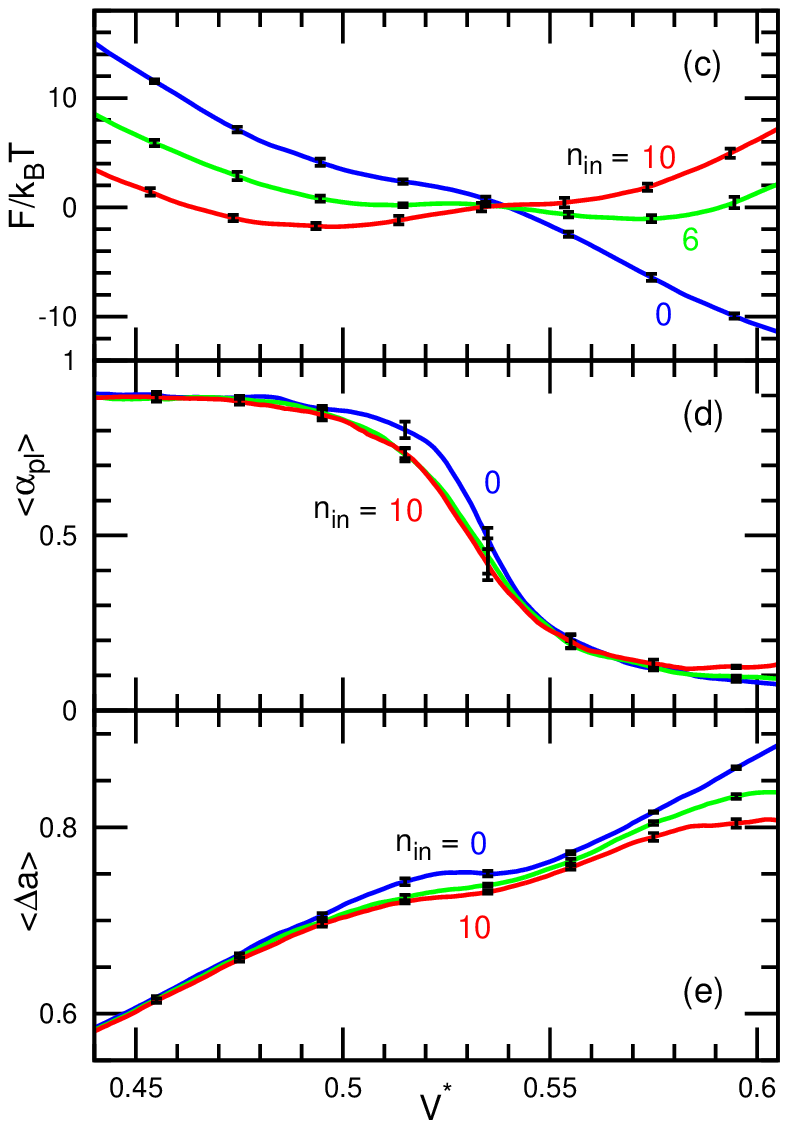}
\caption{
Genus-5 vesicles with $n_{\rm {in}}$ particles in the nucleoplasm for $r_{\rm {po}}/R_{\rm {ves}}=0.2$.
(a),(b) Snapshots at (a) $n_{\rm {in}}=6$ and (b) $n_{\rm {in}}=10$
without the volume constraint.
(c)--(e) Dependence on $V^*$ at $n_{\rm {in}}=0$, $6$, and $10$.
(c) Free-energy profile $F$. (d) Mean aplanarity $\langle\alpha_{\rm {pl}}\rangle$.
(e) Mean area difference $\langle\Delta a\rangle$.
Error bars are shown at several data points.
}
\label{fig:fsp}
\end{figure}

\begin{figure}
\includegraphics{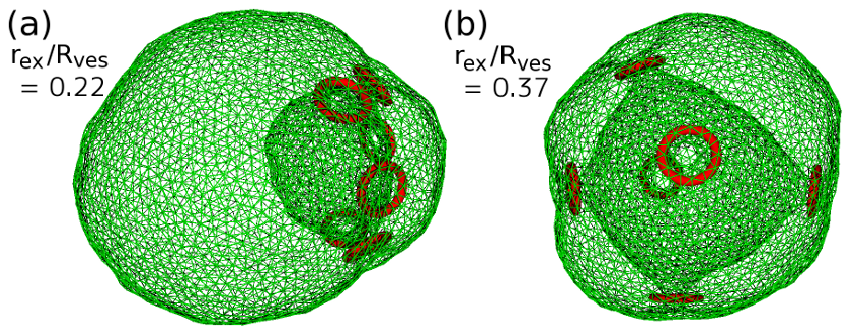}
\includegraphics{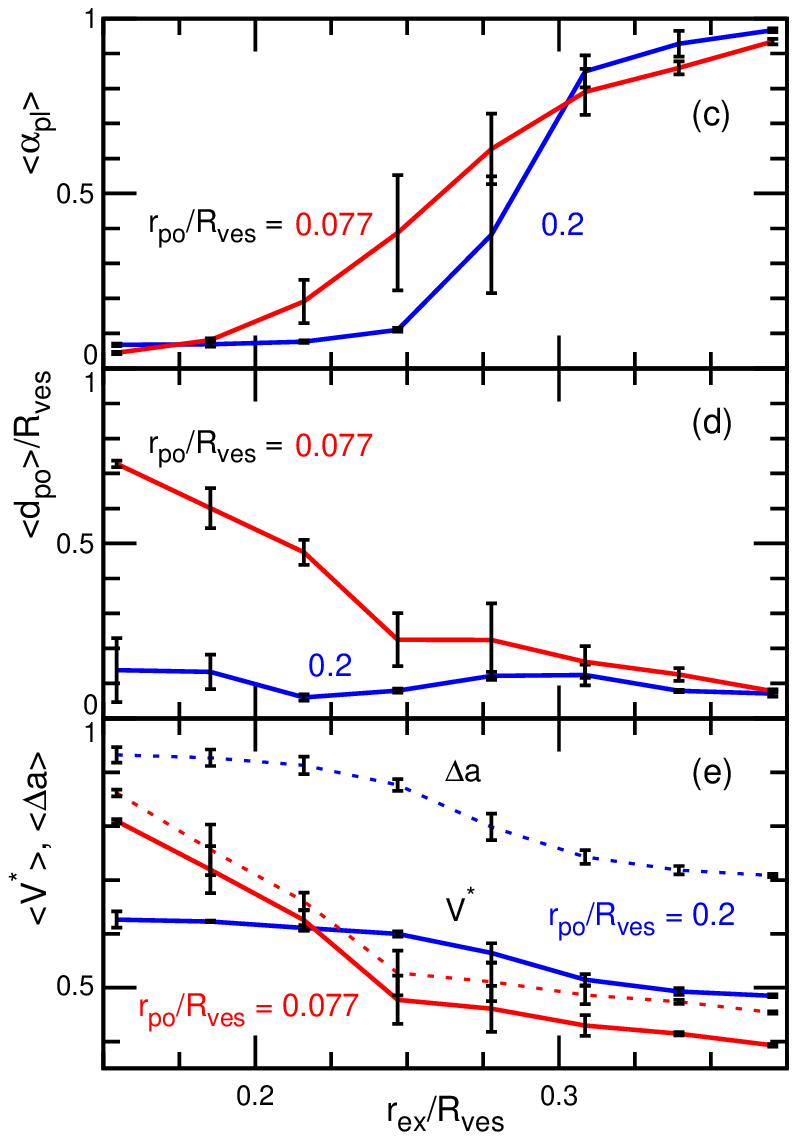}
\caption{
Genus-5 vesicles with repulsion between the ring centers.
(a),(b) Snapshots at (a) $r_{\rm {ex}}/R_{\rm {ves}}=0.22$ and (b) $r_{\rm {ex}}/R_{\rm {ves}}=0.37$
for $r_{\rm {po}}/R_{\rm {ves}}=0.077$.
(c)--(e)  Dependence of the averages of (c) aplanarity $\langle\alpha_{\rm {pl}}\rangle$,
(d) distance $\langle d_{\rm {po}}\rangle$ between 
the centers of the rings and the vesicle, 
(e) reduced volume $\langle V^*\rangle$, and area difference $\langle\Delta a\rangle$.
The solid and dashed lines in (e) represent $\langle V^*\rangle$ and 
$\langle\Delta a\rangle$, respectively.
}
\label{fig:ex}
\end{figure}

As $r_{\rm {po}}$ decreases without the volume constraint,
the oblate bud becomes smaller and moves to the end of the vesicle [see Fig.~\ref{fig:rr}(a)].
As a result, the distance $d_{\rm {po}}$ of the center of $n_{\rm {ring}}$ rings from the center of the vesicle
and $V^*$ both increase as shown in Figs.~\ref{fig:rr}(f) and (g), respectively.
In the smaller bud, the circular pore alignment remains [see Fig.~\ref{fig:rr}(e)].
Thus, the formation of the spherical stomatocyte is not induced alone by the pore-size constraint
either with or without the volume constraint.

Next, in order to find conditions for stabilizing the spherical stomatocyte,
we investigated two additional interactions: 
the osmotic pressure $P_{\rm {in}}$ of the inner bud (nucleoplasm space) and repulsion between the pores.
The pressure $P_{\rm {in}}$ increases with an increase in the number $n_{\rm {in}}$ of inner particles.
The vesicle exhibits the transition to the spherical stomatocyte
with increasing $n_{\rm {in}}$ without a volume constraint,
since the bud volume increases under the transition (see Fig.~\ref{fig:fsp}).
At $n_{\rm {in}}=6$, the circular-cage stomatocyte has $P_{\rm {in}}R_{\rm {ves}}^3/k_{\rm B}T=490 \pm 60$
and is slightly stabler than the spherical stomatocyte,
while the spherical stomatocyte has $P_{\rm {in}}R_{\rm {ves}}^3/k_{\rm B}T=270 \pm 20$  
and is stabler at  $n_{\rm {in}}=10$ [see Figs.~\ref{fig:fsp}(a)--(c)].
Therefore, the pressure induced by ten particles is sufficient to induce the transition.
The transition volume $V^*$ is only slightly reduced by these pressures [see Fig.~\ref{fig:fsp}(d)].

\begin{figure}
\includegraphics{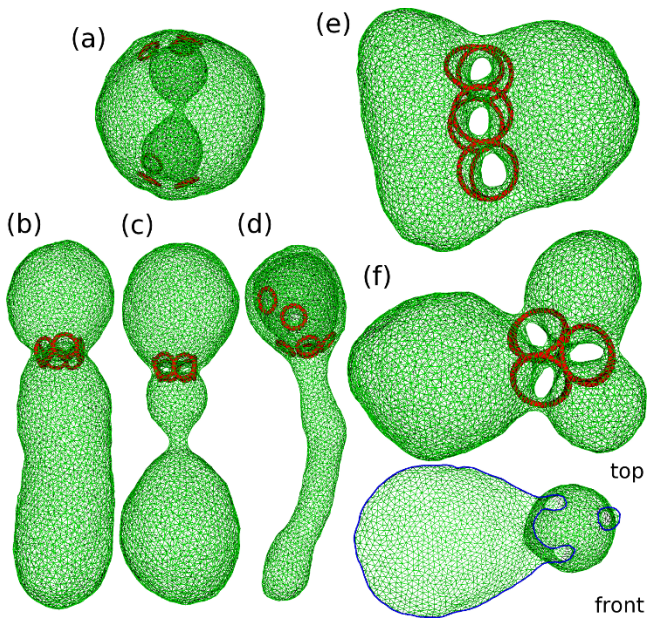}
\caption{
Snapshots of genus-5 vesicles with the ADE potential
at (a)--(d) $r_{\rm {po}}/R_{\rm {ves}}=0.077$ and (e),(f) $r_{\rm {po}}/R_{\rm {ves}}=0.2$.
(a) $V^*=0.6$ and $\Delta a_0=-0.5$.
(b) $V^*=0.6$ and $\Delta a_0=1.5$.
(c) $V^*=0.6$ and $\Delta a_0=2$.
(d) $V^*=0.2$ and $\Delta a_0=1.5$.
(e) $V^*=0.6$ and $\Delta a_0=1.5$.
(f) $V^*=0.6$ and $\Delta a_0=2$.
In the front view of (f),
the front half is removed and the cross sections are indicated by the thick (blue) lines.
}
\label{fig:ade}
\end{figure}

The repulsion between the pores also induces the transition into the spherical stomatocyte
as shown in Fig.~\ref{fig:ex}.
The transition occurs at the excluded radius $r_{\rm {ex}}/R_{\rm {ves}}\simeq 0.25$ and $0.28$
for $r_{\rm {po}}/R_{\rm {ves}}=0.077$ and $0.2$, respectively.
At a large $r_{\rm {ex}}$, the pores are aligned in a pattern of cubic symmetry [see Fig.~\ref{fig:ex}(b)].

Last, we consider the ADE energy.
With the ADE energy, the vesicle can exhibit a variety of shapes as shown in Fig.~\ref{fig:ade}.
As the intrinsic area difference $\Delta a_0$ decreases,
an inner bud of the stomatocyte exhibits a tubular shape and/or budding into two or more compartments [see Fig.~\ref{fig:ade}(a)],
like in genus-0 vesicles.
As $\Delta a_0$ increases, the vesicles form tubular, oblate, and budded shapes,
although the pore constraint prevents the opening of one of the pores for discocyte formation
[see Figs.~\ref{fig:ade}(b)--(f)].
For small rings ($r_{\rm {po}}/R_{\rm {ves}}=0.077$), 
one side of the circular-cage stomatocyte elongates to tubular shapes and subsequently forms spherical buds
at $V^*=0.6$  [see Figs.~\ref{fig:ade}(b),(c)].
At a small $V^*$, a tubular arm elongates from the spherical stomatocyte [see Fig.~\ref{fig:ade}(d)].
Two or three arms can be formed when $\Delta a_0$ or $V^*$ changes rapidly.
The arm radius decreases with a decrease in $V^*$ and an increase in $P_{\rm {in}}$.
For large rings ($r_{\rm {po}}/R_{\rm {ves}}=0.2$), the vesicle forms a discoidal stomatocyte,
in which three pairs of the pores are aligned in a line, at $\Delta a_0=1.5$  [see Fig.~\ref{fig:ade}(e)].
With a further increase in $\Delta a_0$, three buds are formed 
and the pores are located in the branch [see Fig.~\ref{fig:ade}(f)].
The obtained shape transformations are similar to those without the pore-size constraint~\cite{nogu15c},
except for maintaining the inner bud.
We expect that similar shapes are obtained using spontaneous-curvature or bilayer-coupling models~\cite{seif97,svet14},
from the analogy of the morphology of genus-0 vesicles.

Thus, the spherical stomatocyte can be formed
for a small reduced volume $V^*$, a large osmotic pressure $P_{\rm {in}}$, 
and/or large repulsion between the pores.
We will discuss these conditions in more detail in Sec.~\ref{sec:dis}. 

\begin{figure}
\includegraphics{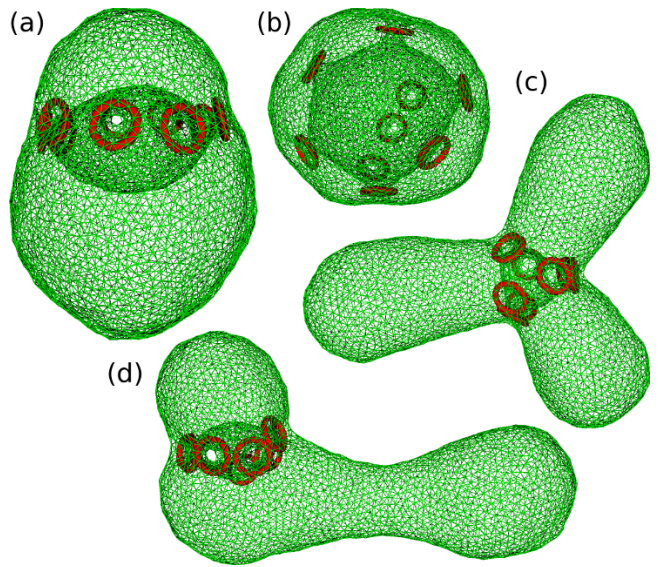}
\caption{
Snapshots of genus-8 vesicles at $r_{\rm {po}}/R_{\rm {ves}}=0.077$
 (a),(b) without and (c),(d) with the ADE potential.
(a) No volume constraint. (b) $V^*=0.4$.
(c), (d) $V^*=0.5$ and $\Delta a_0=1.5$.
}
\label{fig:g8}
\end{figure}

\subsection{Genus-8 vesicles}\label{sec:g8}

To confirm that the above results are not specific to genus-5 vesicles,
we investigated genus-8 vesicles.
Similar vesicle shapes are also obtained for $g=8$ as shown in Fig.~\ref{fig:g8}.
The shape transition from the circular-cage to spherical stomatocytes
occurs with decreasing $V^*$ and the pore-size constraint stabilizes the circular-cage shape.
For the same ring radius, the location of the inner bud 
deviates less from the vesicle center
[see Fig.~\ref{fig:g8}(a)].
The pressure $P_{\rm {in}}$ induces the transition into the spherical stomatocyte.
A large $\Delta a_0$ induces tubular and budded stomatocytes.
Elongated circular-cage to spherical stomatocyte and three-armed vesicles can coexist
as (meta-)stable states as shown in Figs.~\ref{fig:g8}(c) and (d).

\section{Discussion}\label{sec:dis}

We have investigated vesicle morphology under pore-size constraint.
It is found that the pore-size constraint suppresses the transition from the circular-cage to spherical stomatocytes. 
However, the spherical stomatocytes can be stabilized
by the following additional interactions:
reduction of the perinuclear volume, an increase in nucleoplasm volume produced by the osmotic pressure, 
and repulsion between the pores.

Let us consider these conditions for the spherical stomatocyte shape on the nuclear envelope.
Our simulation shows that the osmotic pressure generated by only ten particles is needed
to stabilize the spherical stomatocyte.
The nucleoplasm space is densely filled with nucleosomes, nuclear lamina, and other proteins,
and so a much larger pressure is expected.
Thus, this osmotic pressure seems to be sufficient but we also discuss the other conditions.
Although the NPCs do not directly interact with each other,
they can interact indirectly via the lamina and other proteins. 
Hence, it is possible that an effective repulsion exists between them
and it might help the spherical-stomatocyte formation.

Since the nuclear envelope is connected to the ER, 
its perinuclear volume and area difference are shared with the ER.
The ER has complicated structures including planar and tubular networks \cite{shib09,tera13,west15}
so that the total volume and area difference are rather determined by the ER.
Therefore, they are not likely to be control parameters for the nuclear shape.
At a large area difference, a few arms can elongate from the stomatocyte.
However, the arms exhibit only tubular or budded shapes. Hence,
a planar membrane and linear or helicoidal edges cannot be formed by the present membrane system.
Thus, the local curvature regulation is likely necessary to construct the ER shapes.
In living cells, Bin-Amphiphysin-Rvs (BAR) superfamily proteins and other proteins participate to
control local membrane curvatures in the ER and other organelles \cite{mcma05,shib09,baum11,itoh06,masu10,mim12a,simu15a}.
Tubulation from a vesicle or flat membrane and the stabilization of round edges of disk-shaped vesicles have been simulated
with the inclusions of an anisotropic spontaneous curvature like the BAR proteins \cite{simu15a,rama13,nogu14,nogu16}.
To reproduce the whole shape of the nucleus and ER, taking into account such anisotropic inclusions 
is one of the important extensions that can be explored in further studies.

Since the nucleus is a highly complex organelle,
we could not incorporate all of the interactions that the nuclear envelope has with other cellular entities into our model in this study.
We briefly discuss some other factors that may modify its shape.
(i) Membrane asymmetry:
Here, a homogeneous membrane is considered.
However, it is known that the outer and inner bilayers of the nuclear envelope have different compositions of lipids and proteins.
The outer and inner monolayers of each bilayer are also different.
Such asymmetric compositions can induce spontaneous curvatures.
(ii) Protein interactions:
The inner bilayer contacts the lamina that interact with the nucleosomes.
We consider only an isotropic pressure here, but modeling the inhomogeneous stress induced by the lamina and other proteins 
is also important to understand abnormal nuclear shapes in disease states.
Protein bridges between the outer and inner bilayers may maintain the distance between the two bilayers.
Here, we consider that the pore-size constraint effect of the NPC but the NPC binding to the membrane also changes local membrane curvature
and influences the pore shape.
Thus, we cannot completely exclude these other possibilities.
However,  our results clarified that the shape of the nuclear envelope, i.e., a spherical stomatocyte, can be easily formed in several ways,
and we concluded that it is a very robust structure. 

Last, let us briefly discuss possible applications of our pore-size constraint approach.
Neck structures like a nuclear pore can be found in other organelles.
For example, the inner membrane of mitochondrion has numerous invaginations called cristae \cite{mann06,sche08}.
It is known that the planer cristae are connected to the inner membrane via a narrow tubular neck.
One may use a toroidal ring to restrict this neck radius as a protein model and investigate the cristae structures.
Thus, the morphological analysis of other membrane structures involving narrow necks are interesting topics for the further studies.

\begin{acknowledgments}
This work was supported by JSPS KAKENHI Grant Number JP25103010.
\end{acknowledgments}

\end{document}